\documentstyle[pra,aps,eqsecnum,epsf,array]{revtex}

\newcommand{\be}{\begin{equation}}
\newcommand{\ee}{\end{equation}}
\newcommand{\ben}{\begin{eqnarray}}
\newcommand{\een}{\end{eqnarray}}

\begin{document}

\twocolumn[\hsize\textwidth\columnwidth\hsize\csname 
@twocolumnfalse\endcsname 
 
\title{Anharmonic oscillator radiation process in a large cavity} 
\author{G. Flores-Hidalgo$^{\star}$, 
A. P. C. Malbouisson$^{\star}$}

\address{$^{\star}\;$Centro Brasileiro de Pesquisas F\'\i sicas - CBPF/MCT, 
Rua Dr. Xavier Sigaud 150\\
22290-180, Rio de Janeiro, RJ, Brazil} 
\date{\today} 
 
\maketitle 
 
\begin{abstract}
We consider a particle represented by  an anharmonic oscillator, 
coupled to an environment (a field) modeled by 
an ensemble of anharmonic oscillators, the whole system being confined 
in a cavity of diameter $L$. Up to the first perturbative order in the 
quartic interaction (interaction parameter $\lambda$), we use the formalism  
of {\it dressed} states introduced in previous publications, to obtain for a 
large cavity explicit $\lambda$-dependent formulas for the particle radiation 
process. These formulas are obtained in terms of the corresponding exact 
expressions for the linear case. We conclude for the enhancement of the 
particle decay induced by the quartic interaction.\\
\vspace{0.34cm}
\noindent
PACS Number(s):~03.65.Ca, 32.80.Pj
\end{abstract} 
\vskip2pc]

\newpage

\section{Introduction}

In previous publications
\cite{adolfo1,adolfo2,gabriel}, a non perturbative approach ({\it dressed} 
states) has been used to 
study systems that can be 
described by a Hamiltonian of the form,
\begin{equation} 
H=\frac{1}{2}\left[p_{0}^{2}+\omega_{0}^{2}q_{0}^{2}+ 
\sum_{k=1}^{N}(p_{k}^{2}+\omega_{k}^{2}q_{k}^{2})\right]-q_{0}
\sum_{k=1}^{N}c_{k}q_{k},  
\label{Hamiltoniana} 
\end{equation} 
where the limit $N\rightarrow\infty$ is understood, the subscript $0$ refers to 
a particle approximated by a harmonic oscillator having {\it bare} frequency 
$\omega_{0}$, and $k=1,2,...N$ refer to the harmonic field modes. A 
Hamiltonian of the type of Eq.(\ref{Hamiltoniana}), can be viewed as a linear 
coupling of an atom with the scalar potential, and has been investigated in 
\cite{adolfo1,adolfo2} to define rigorous {\it dressed} states which allow 
a non-perturbative approach to the time evolution of atomic states. An 
Hamiltonian of the type (\ref{Hamiltoniana}) has also been employed in 
\cite {paz} to study the quantum Brownian motion of a particle with the 
path-integral formalism. In the limit $N\rightarrow \infty$ we recover, 
within the framework of the harmonic approximation, the situation of the 
linear coupling of an atom with the scalar potential, or the coupling 
of a Brownian particle with its environment, after redefinition of 
divergent quantities. In the case of the coupled atom field system, the 
above mentioned formalism of {\it dressed} states recovers the experimental 
observation that excited states of atoms in sufficiently small cavities 
are stable. It allows to give formulas for the probability of an atom to 
remain excited for an infinitely long time, provided it is placed in a 
cavity of appropriate size \cite{adolfo2}.

In this note we intend to generalize, making some appropriate approximations, 
to an {\it anharmonic} oscillator, the linear coupling to an environnement as 
it has been considered in the above mentioned works. In this case, the whole 
system is described by the Hamiltonian,
\be
H_{1}=
H(p_{0},q_{0},\{p_{k},q_{k}\})+\sum_{r=0}^N\lambda_{r}{\cal T}^{(r)}_{\mu\nu\rho\sigma}
q_{\mu}q_{\nu}q_{\rho}q_{\sigma},
\label{H1}
\ee 
where $H(p_{0},q_{0},\{p_{k},q_{k}\})$ is the bilinear Hamiltonian in Eq.
(\ref{Hamiltoniana}) and $T^{(r)}_{\mu\nu\rho\sigma}$ are some coefficients 
that will be defined below. In Eq. (\ref{H1}) summation over repeated greek 
labels is understood. We emphasize that our problem is different from the 
situation treated in the pioneering papers of Refs\cite{bender,wu}. We do not 
intend to go to higher orders in the perturbative series for the energy 
eigenstatates, we will remain at a first order correction in $\lambda_r$, 
and we will try to see what are the effects of the anharmonicity term, given 
by the last term of Eq. (\ref{H1}), on our previous results for the linear 
coupling with an environnement. We  notice that the anharmonicity term in 
Eq. (\ref{H1}) involves, independent quartic terms of the type $q_\mu^4$, 
$\mu=0,\{i\}$ (self-coupling of the bare oscillator and of the field modes), 
quartic terms coupling the oscillator to the field modes and also  the terms 
coupling the field modes among themselves. These terms are of the type 
$q_0q_i^3$, $q_0^2q_i^2$, $q_0^3q_i$, and of the type $q_i^nq_j^m$, with 
$n+m=4$, for the coupling between the field modes. We intend to start from 
the exact solutions we have found in the linear case, and investigate at 
first order, how the quartic interaction characteristic of the anharmonicity 
changes the decay probabilities obtained in the linear case.

In the section {\bf II} we review some results obtained in the previous 
publications mentioned above. In section {\bf III} we show how, making 
appropriate approximations, the {\it dressed} states approach introduced for 
ohmic systems can be used to generalize some results to an anharmonic 
oscillator coupled to an ohmic environnment.
    
\section{The harmonic system}

The bilinear Hamiltonian (\ref{Hamiltoniana}) can be turned to principal axis 
by means of a point transformation, 
$q_{\mu}=t_{\mu}^{r}Q_{r},~p_{\mu}=t_{\mu}^{r}P_{r};~ 
\mu=(0,\{k\}),\;k=1,2,..., N;~ r=0,...N,$ 
performed by an orthonormal  matrix $T=(t_{\mu}^{r})$. The subscript $\mu=0$  
refers to the atom (or the Brownian particle) and $\mu=k\;, k=1,2,3...$ 
refer to the harmonic modes  of the field (or the thermal bath) The 
subscripts $r$ refer to the normal modes. In terms of normal momenta and 
coordinates, the transformed Hamiltonian in principal axis reads,
\be
H=\frac{1}{2}\sum_{r=0}^{N}(P_{r}^{2}+\Omega_{r}^{2}Q_{r}^{2}),
\label{eixospr.}
\ee
where the $\Omega_{r}$'s are the normal frequencies corresponding to the 
possible collective stable oscillation modes of the coupled system. The 
matrix elements $t_{\mu}^{r}$ are given by \cite{adolfo1}
\begin{equation} 
t_{k}^{r}=\frac{c_{k}}{(\omega_{k}^{2}-\Omega_{r}^{2})}t_{0}^{r}\;,\;\; 
t_{0}^{r}= \left[1+\sum_{k=1}^{N}\frac{c_{k}^{2}}{ 
(\omega_{k}^{2}-\Omega_{r}^{2})^{2}}\right]^{-\frac{1}{2}}
\label{tkrg1} 
\end{equation} 
with the condition,
\begin{equation} 
\omega_{0}^{2}-\Omega_{r}^{2}=\sum_{k=1}^{N}\frac{c_{k}^{2}}
{\omega_{k}^{2} -\Omega_{r}^{2}}.  
\label{Nelson1} 
\end{equation}

We take $c_{k}=\eta (\omega_{k})^{n}$. In this case the environment is 
classified according to $n>1$, $n=1$, or $n<1$, respectively as 
{\it supraohmic}, {\it ohmic} or {\it subohmic}. For a subohmic environment 
the sum in Eq.(\ref{Nelson1}) is convergent and the frequency $\omega_{0}$ 
is well defined. For ohmic and supraohmic environments the sum in the right 
hand side of Eq.(\ref{Nelson1}) diverges what makes the equation meaningless 
as it stands, a renormalization procedure being needed. We restrict ourselves 
to {\it ohmic} systems. In this case, using the method described in 
\cite{adolfo1} we can define a {\it renormalized} frequency $\bar{\omega}$, by 
means of a counterterm $\delta \omega^{2}$,
\be
\bar{\omega}^{2}=\omega_{0}^{2}-\delta \omega^{2}\;;\;\;\;\;\delta \omega^{2}=N\eta^{2},
\label{omegabarra}
\ee
in terms of which Eq.(\ref{Nelson1}) becomes,
\be
 \bar{\omega}^{2}-\Omega_{r}^{2}=\eta^{2}\sum_{k=1}^{N}\frac{\Omega_{r}^{2}}
{\omega_{k}^{2} -\Omega_{r}^{2}},  
\label{Nelson3} 
\ee
We see that in the limit $N\rightarrow \infty$ the above procedure is exactly 
the analogous of naive mass renormalization in Quantum Field Theory: the 
addition of a counterterm 
$-\delta \omega^{2}q_{0}^{2}$ allows to compensate the infinuty of 
$\omega_{0}^{2}$ in such a way as to leave a finite, physically meaninful 
renormalized frequency $\bar{\omega}$. This simple renormalization scheme 
has been originally introduced in Ref.\cite{Thirring}.

To proceed, we take the constant $\eta$ as $\eta=\sqrt{2g\Delta\omega}$, 
$\Delta\omega$ being the interval between two neighbouring bath frequencies 
(supposed uniform) and where $g$ is some constant (with dimension of 
$frequency$). We restrict ouselves to the physical situations in which the 
whole system is confined to a cavity of diameter $L$, in which case the 
environment (field) frequencies $\omega_k$ can be writen in the form 
\be
\omega_k=2k\pi/L,\;\;\;\;k=1,2,...\;.
\label{discreto}
\ee
Then using the formula, 
\begin{equation}
\sum_{k=1}^{N}\frac{1}{(k^{2}-u^{2})}= \left[\frac{1}{2u^{2}}-\frac{\pi}{u}
{\rm cot}(\pi u)\right],
\label{id4}
\end{equation}
and restricting ourselves to an {\it ohmic} environment, Eq.(\ref{Nelson3}) 
can be written in closed form,
\begin{equation} 
{\mathrm{cot}}(\frac{L\Omega}{2c})=\frac{\Omega}{\pi g}+\frac{c}
{L\Omega}(1-\frac{\bar{\omega}^{2}L}{\pi gc}).  
\label{eigenfrequencies1} 
\end{equation}
The solutions of Eq.(\ref{eigenfrequencies1}) with respect to  $\Omega$ give 
the spectrum of eigenfrequencies $\Omega_{r}$ corresponding to the collective 
normal modes. The transformation matrix elements turning the  system to 
principal axis are obtained in terms of the physically meaningful quantities 
$\Omega_{r}$, $\bar{\omega}$, after some rather long but straightforward 
manipulations analogous as it has been done in \cite{adolfo1}. They read,  
\begin{eqnarray} 
t_{0}^{r}&=&\frac{\eta \Omega_{r}}{\sqrt{(\Omega_{r}^{2}-\bar{ 
\omega}^{2})^{2}+\frac{\eta^{2}}{2}(3\Omega_{r}^{2}-\bar{\omega}^{2})+
\pi^{2}g^{2}\Omega_{r}^{2}}}\;, \nonumber \\   
& & t_{k}^{r}=\frac{\eta\omega_{k}}{\omega_{k}^{2}-\Omega_{r}^{2}}t_{0}^{r}. 
\label{t0r2} 
\end{eqnarray}

To study the time evolution of the system, we start from the eigenstates of 
our system, $|\left. n_{0},n_{1},n_{2}...\right>$, represented by the 
normalized eigenfunctions in terms of the normal coordinates $\{Q_{r}\}$,  
\begin{eqnarray} 
\phi_{n_{0}n_{1}n_{2}...}(Q,t)&=&\prod_{s}\left[\sqrt{\frac{2^{n_s}}{n_s!}}H_{n_{s}}
(\sqrt{\frac{ 
\Omega_{s}}{\hbar}}Q_{s})\right]\times\nonumber\\
& &~~~~~~\Gamma_{0}e^{-i\sum_{s}n_{s}\Omega_{s}t}, 
\label{autofuncoes} 
\end{eqnarray} 
where $H_{n_{s}}$ stands for the $n_{s}$-th Hermite polynomial and 
$\Gamma_{0}$ is the normalized vacuum eigenfunction. We introduce 
{\it dressed} coordinates $q^{\prime}_{0}$ and $ \{q^{\prime}_{i}\}$ for, 
respectively the {\it dressed} atom and the {\it dressed} modes of the field, 
defined by \cite{adolfo1},  
\begin{equation} 
\sqrt{\frac{\bar{\omega}_{\mu}}{\hbar}}q^{\prime}_{\mu}=\sum_{r}t_{\mu}^{r} 
\sqrt{\frac{\Omega_{r}}{\hbar}}Q_{r},  
\label{qvestidas1} 
\end{equation} 
valid for arbitrary $L$ and where $\bar{\omega}_{\mu}=\{\bar{\omega}, 
\;\omega_{i}\}$. In terms of the bare coordinates the dressed coordinates 
are expressed as, 
\begin{equation}
q^{\prime}_{\mu}=\sum_{\nu}\alpha_{\mu \nu}q_{\nu},  
\label{qvestidas3}
\end{equation}
where 
\begin{equation}
\alpha_{\mu \nu}=\frac{1}{\sqrt{\bar{\omega}_{\mu}}}\sum_{r}t_{\mu}^{r}t_{%
\nu}^{r}\sqrt{\Omega_{r}}.  
\label{qvestidas4}
\end{equation}

In terms of the {\it dressed} coordinates, we  define for a fixed instant 
{\it dressed} states, $|\left. \kappa_{0},\kappa_{1},\kappa_{2}...\right>$ 
by means of the complete orthonormal set of functions \cite{adolfo1},  
\begin{equation} 
\psi_{\kappa_{0} \kappa_{1}...}(q^{\prime})=\prod_{\mu}\left[\sqrt{
\frac{2^{\kappa_{\mu}}}{\kappa_{\mu}!}} 
H_{\kappa_{\mu}} (\sqrt{\frac{\bar{\omega}_{\mu}}{\hbar}} 
q^{\prime}_{\mu})\right]\Gamma_{0},  
\label{ortovestidas1} 
\end{equation} 
where $q^{\prime}_{\mu}=q^{\prime}_{0},\, q^{\prime}_{i}$, $\bar{\omega} 
_{\mu}=\{\bar{\omega},\, \omega_{i}\}$. Note that the ground state 
$\Gamma_{0}$ in the above equation is the same as in Eq.(\ref{autofuncoes}). 
The invariance of the ground state is due to our definition of {\it dressed} 
coordinates given by Eq. (\ref{qvestidas1}). Each function 
$\psi_{\kappa_{0} \kappa_{1}...}(q^{\prime})$ describes a state in which 
the {\it dressed} oscillator $q'_{\mu}$ is in its $\kappa_{\mu}-th$ excited 
state. Let us consider the {\it dressed} state 
$|\left.0,0,...1(\mu),0...\right>$, represented by the wavefunction 
$\psi_{00...1(\mu)0...}(q')$. It describes the configuration in which only 
the dressed oscillator $q^{\prime}_{\mu}$  is in the first excited level. 
Then it is shown in \cite{adolfo1} the following expression for the time 
evolution of the first-level excited dressed oscillator $q^{\prime}_{\mu}$,  
\be 
|\left.0,0,...1(\mu),0...\right>(t)=
\sum_{\nu}f^{\mu \nu}(t)|\left.0,0,...1(\nu),0...\right>(0)\;,
\label{ortovestidas5}
\ee
where
\be
f^{\mu \nu}(t)=\sum_{s}t_{\mu}^{s}t_{\nu}^{s}e^{-i\Omega_{s}t}.
\label{ortovestidas6} 
\ee
From Eq.(\ref{ortovestidas5}) we see that the initially excited dressed 
oscillator naturally distributes its energy among itself and all other 
dressed oscillators (the atom and the environment) as time goes on, with 
probability amplitudes given by the quantities $f^{\mu \nu}(t)$ in Eq.
(\ref{ortovestidas6}). For $\mu=0$ in Eq. (\ref{ortovestidas5}) the 
coefficients $f^{0 \nu}(t)$ have a simple interpretation: $f^{00}(t)$ 
and $f^{0i}(t)$  are respectively the probability amplitudes that at time 
$t$ the dressed particle still be excited or have radiated a quantum of 
frequency $\hbar \omega_{i}$. We see that this formalism allows a quite 
natural description of the radiation process as a simple exact time 
evolution of the system. In the case of a very large cavity (free space) our 
method reproduces for weak coupling the well-known perturbative results 
\cite{adolfo1,adolfo2}.
\section{An anharmonic oscillator in a large cavity}

The introduction of the quartic interaction term in Eq.(\ref{H1})
changes the Hamiltonian in principal axis from Eq.(\ref{eixospr.}) into
\begin{eqnarray}
H_{1}&=&\frac{1}{2}\sum_{r=0}^{N}(P_{r}^{2}+\Omega_{r}^{2}Q_{r}^{2})+\nonumber\\
& &~~~~~~~~~\sum_{r=0}^N\lambda_r{\cal T}^{(r)}_{\mu\nu\rho\sigma}
t_\mu^{r_1}t_\nu^{r_2}t_\rho^{r_3}t_\sigma^{r_4}
Q_{r_1}Q_{r_2}Q_{r_3}Q_{r_4}\;,
\label{eixospri.2}
\end{eqnarray}
where summation over the repeated indices $r_{1},r_{2},r_{3},r_{4}$ is 
understood. In order to have a specific quartic interaction, we make a 
choice for the coefficients ${\cal T}_{\mu\nu\rho\sigma}^{(r)}$ in the above 
equation,
\be
{\cal T}^{(r)}_{\mu\nu\rho\sigma}=t_{\mu}^rt_{\nu}^rt_{\rho}^rt_{\sigma}^r\;,
\label{T1}
\ee
which replaced in Eq. (\ref{eixospri.2}) and using the orthonormality of the 
matrix $t_\mu^r$ gives the Hamiltonian in principal axis,
\be
H_{1}=\frac{1}{2}\sum_{r=0}^{N}(P_{r}^{2}+\Omega_{r}^{2}Q_{r}^{2})
+\sum_{r=0}^N\lambda_rQ_r^4\;.
\label{H2}
\ee

Since the quartic interaction, as given by Eq. (\ref{T1}),  decouples the
normal coordinates, the renormalization procedure will remains the same as in
the absence of the quartic interaction. This means that the dressed frequency
$\bar{\omega}$ is still given by Eq. (\ref{omegabarra}).  

Performing a perturbative calculation in $\lambda_r$, we can obtain the 
first order correction to the energy of the system, 
$\sum_{r}\hbar \Omega_{r}$, in such a way that the $\lambda_r$-corrected 
energy can be written in the form,
\be
E(\{\lambda_r\})=\hbar\sum_{r}\left(\Omega_{r}+\lambda_r e_{r}^{(1)}\right)\;,
\label{energ.1}
\ee
where the first order correction $e_{r}^{(1)}$ is given by 
\be
e_{r}^{(1)}=\frac{15}{4\Omega_r^2}\;.
\label{first}
\ee

For sufficiently small $\lambda_r$ we will have quasi-harmonic normal 
collective modes having frequencies $\Omega_{r}+\lambda_r e_{r}^{(1)}$. 
Accordingly we can describe {\it approximatelly} the system in terms of 
modified harmonic eigenstates, which can be written as a generalization of 
the exact eigenstates (\ref{autofuncoes}), replacing the eigenfrequencies 
$\Omega_{r}$ by the $\lambda$-corrected values 
$\Omega_{r}+\lambda_{r}e_{r}^{(1)}$,
\begin{eqnarray} 
\phi_{n_{0}n_{1}n_{2}...}(Q,t;\{\lambda_{r}\})&=&
\prod_{s}\left[\sqrt{\frac{2^{n_s}}{n_s!}}H_{n_{s}}(\sqrt{\frac{ 
\Omega_{s}}{\hbar}}Q_{s})\right]\times\nonumber\\
& &~~~~~~
\Gamma_{0}e^{-i\sum_{s}n_{s}\left(\Omega_{s}+\lambda_r e_{r}^{(1)}\right)t}. 
\label{autofuncoes1} 
\end{eqnarray}
From the modified harmonic eigenstates (\ref{autofuncoes1}), we can follow 
analogous steps as in the harmonic case \cite{adolfo1,adolfo2} to study the 
$\lambda$-corrected evolution of a dressed particle, generalizing Eq.
(\ref{ortovestidas5}),   
\begin{eqnarray}
|\left.1,0,0,...,0...\right>(t;\{\lambda_s\})&=&\sum_{\nu}f^{0\nu}(t;\{\lambda_s\})
\nonumber\\
& &~\times|\left.0,0,...1(\nu),0...\right>(0)\;,
\label{ortovestidas7} 
\end{eqnarray}
where
\be
f^{0 \nu}(t;\{\lambda_s\})=
\sum_{s}t_{0}^{s}t_{\nu}^{s}e^{-i\left(\Omega_{s}+\lambda_s e_{s}^{(1)}\right)t}
\label{ortovestidas8} 
\ee

For a very large cavity, from Eq. (\ref{t0r2}) we obtain for $L$ arbitrarily 
large, 
\begin{equation}
t_{0}^{r}\rightarrow Lim_{L \rightarrow \infty}\frac{\sqrt{2g}
\Omega \sqrt{2\pi c/L}}{\sqrt{(\Omega^{2}-\bar{\omega}
^{2})^{2}+\pi^{2}g^{2}\Omega^{2}}}.  
\label{t0r(R)}
\end{equation}
from which using the definition of the coefficients 
$f^{0\nu}(t;\{\lambda_s\})$ from Eq. (\ref{ortovestidas8}), and the fact that 
for very large values of $L$, $ 2\pi c/L=\delta \omega = \delta \Omega$, we 
have an expression for the $\{\lambda_r\}$-corrected probability amplitude 
for the particle be still excited after an ellapsed time $t$, the quantity,
\begin{equation}
f^{00}(t;\{\lambda_\Omega\})=\int_{0}^{\infty}\frac{2g\Omega^{2}e^{-i\left(\Omega+
\frac{15}{4\Omega^2}\lambda_\Omega\right) t}\; d\Omega} 
{(\Omega^{2}-\bar{\omega}^{2})^{2}+\pi^{2}g^{2}\Omega^{2}}.  
\label{f00(1)}
\end{equation}
From dimensional arguments, we can choose $\lambda_\Omega=\lambda\Omega^3$, 
where $\lambda$ is a dimensionless small fixed constant. With this choice, 
after expanding inpowers of $\lambda$ the exponential in Eq. (\ref{f00(1)}), 
we obtain to first order in $\lambda$ the amplitude,
\be
f^{00}(t;\lambda)=f^{00}(t)+\frac{15\lambda t}{4}\frac{\partial}{\partial t} 
f^{00}(t)\;,
\label{f00(2)}
\ee
where $f^{00}(t)$ is the probability amplitude for the harmonic problem, that 
the particle be still excited after a time $t$ \cite{adolfo1,adolfo2}. For 
$\frac{\pi g}{2}< \bar{\omega}$ (situation that includes weak coupling, 
$g\ll \bar{\omega}$) and for a very large cavity, $f^{00}(t)$ is given by
\cite{gabriel},
\be
f^{00}(t)=\left(1-\frac{i\pi g}{2\kappa}\right)
e^{-i\kappa t-\pi gt/2}+iJ(t),
\label{f00(3)}
\ee
where, 
\begin{eqnarray}
J(t)&=& 2g\int_0^{\infty}dy\frac{y^2 e^{-yt}}{(y^2+\bar{\omega}^2)^2-
\pi^2g^2y^2}\approx 
\nonumber \\
& & \approx \frac{4g}{\bar{\omega}^4t^3}\;\;\;(t\gg \frac{1}{\bar{\omega}})
\label{J}
\end{eqnarray}

From Eq. (\ref{f00(2)}) we can obtain at order $\lambda$, the probability 
that the particle remains in the first excited state at time $t$,
\be
|f^{00}(t;\lambda)|^2=|f^{00}(t)|^2+\frac{15\lambda t}{4}\frac{\partial}{\partial t}
|f^{00}(t)|^2\;.
\label{f00(4)}
\ee
We know that $|f^{00}(t)|^2$ is a decreasing function of $t$, what means that
the derivative of this function with respect to $t$ is negative. Therefore, 
since $\lambda>0$, we conclude from Eq. (\ref{f00(4)}) that 
$|f^{00}(t;\lambda)|^2$ is smaller than the harmonic probability
$|f^{00}(t)|^2$. Indeed we know from Refs. \cite{adolfo2,gabriel} that for 
large $t$ ($t>> \frac{1}{\bar{\omega}}$) we have,
\begin{eqnarray} 
|f^{00}(t)|^{2}&=&(1+\frac{\pi^{2}g^{2}}{4\kappa^{2}})e^{-\pi gt}-
\frac{8 g}{ \bar{\omega}^{4}t^{3}}[\sin(\kappa t)
+\nonumber\\
& &~~~~~~\frac{\pi g}{2\kappa} \cos(\kappa t)]e^{-\pi gt/2}+
\frac{ 16g^{2}}{\bar{\omega}^{8}t^{6}}\;,   
\label{f00(5)} 
\end{eqnarray} 
which is a rapidly decreasing function of $t$. 

In $Fig.1$ we plot on the same scale the $\lambda$-corrected probability 
(\ref{f00(4)}) and the harmonic probability in Eq.(\ref{f00(5)}), for 
$\bar{\omega}=4.0\times 10^{14}/s$ and $g=\alpha \bar{\omega}$, where 
$\alpha$ is the fine structure constant, $\alpha=1/137$ and time is rescaled 
as  $t\times 10^{-13}s$. The solid line is the harmonic probability 
(\ref{f00(5)}) and the dashed line is the $\lambda$-corrected probability 
(\ref{f00(4)}), for $\lambda=1/50$. We see clearly the enhancement of the 
particle decay induced by the quartic interaction.

\begin{figure}[c] 
\epsfysize=6cm  
{\centerline{\epsfbox{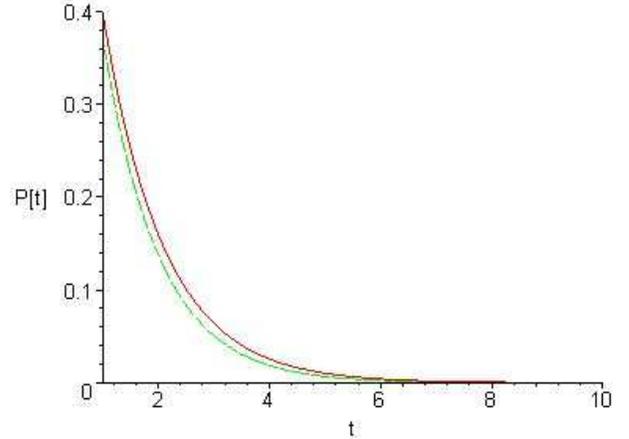}}}
\caption{Plot on the vertical axis, 
commonly named $P[t]$, of the $\lambda$-corrected probability (\ref{f00(4)}) 
(dashed line) and the harmonic probability  Eq.(\ref{f00(5)}) (solid line), 
for $\bar{\omega}=4.0\times 10^{14}/s$ and $g=\alpha \bar{\omega}$. $\alpha$ 
is the fine structure constant, $\alpha=1/137$ and time in units of 
$10^{-13}s$}
 
\end{figure}

\section{Acknowledgements}
This work has been partially supported by CNPq (Brazilian National 
Research Council)

\end{document}